\documentclass[10pt,amsmath,amssymb,aps,nofootinbib,notitlepage,prd,superscriptaddress,twocolumn]{revtex4-1}

\let\cite\citep

\usepackage{lineno}

\usepackage{xspace}
\usepackage[T1]{fontenc}
\usepackage{aas_macros,color,graphicx,hyperref,mathrsfs,orcidlink}

\usepackage{newpxtext,newpxmath}

\definecolor{fed_blue}{HTML}{07004D}

\definecolor{steel_blue}{HTML}{2D82B7}
\definecolor{steel_blue_dark}{HTML}{1C71A6}
\definecolor{aqua_marine}{HTML}{42E2B8}
\definecolor{dutch_white}{HTML}{F3DFBF}
\definecolor{light_coral}{HTML}{EB8A90}
\definecolor{light_coral_dark}{HTML}{BA5A60}

\hypersetup{
    colorlinks = true,
    linkcolor = purple,
    urlcolor  = steel_blue_dark,
    citecolor = steel_blue_dark,
    anchorcolor = black
}
\newcommand{\appref}[1]{\hyperref[#1]{Appendix~\ref*{#1}}}

\let\originalleft\left
\let\originalright\right
\renewcommand{\left}{\mathopen{}\mathclose\bgroup\originalleft}
\renewcommand{\right}{\aftergroup\egroup\originalright}

\newcommand{\fracsys}{f_{\textrm{sys}}\xspace}
\newcommand{\krangjl}{\texttt{Krang.jl}\xspace}
\newcommand{\SNR}{$S/N$\xspace}
\renewcommand{\vec}[1]{\mathbf{#1}}
\newcommand{\sigmastar}{\sigma_{a_\ast}}

\defcitealias{Thompson2001}{TMS}

\begin{document}

\title{\textbf{Spin inference with the Black Hole Explorer. I. Fisher information matrix forecast}}

\author{Joseph R. Farah\,\orcidlink{0000-0003-4914-5625}}
\email{josephfarah@ucsb.edu}
\affiliation{Center for Astrophysics $|$ Harvard \& Smithsonian, 60 Garden St, Cambridge, MA 02138, USA}
\affiliation{Black Hole Initiative, Harvard University, 20 Garden St, Cambridge, MA 02138, USA}
\affiliation{Miller Institute for Basic Research in Science, 206B Stanley Hall, Berkeley, CA 94720, USA}
\affiliation{Department of Astronomy, University of California, Berkeley, CA 94720-3411, USA}

\author{Daniel C. M. Palumbo}
\affiliation{Center for Astrophysics $|$ Harvard \& Smithsonian, 60 Garden St, Cambridge, MA 02138, USA}
\affiliation{Black Hole Initiative, Harvard University, 20 Garden St, Cambridge, MA 02138, USA}

\author{Michael D. Johnson\,\orcidlink{0000-0002-4120-3029}}
\affiliation{Center for Astrophysics $|$ Harvard \& Smithsonian, 60 Garden St, Cambridge, MA 02138, USA}
\affiliation{Black Hole Initiative, Harvard University, 20 Garden St, Cambridge, MA 02138, USA}

\author{Dominic O. Chang\,\orcidlink{0000-0001-9939-5257}}
\affiliation{Center for Astrophysics $|$ Harvard \& Smithsonian, 60 Garden St, Cambridge, MA 02138, USA}
\affiliation{Black Hole Initiative, Harvard University, 20 Garden St, Cambridge, MA 02138, USA}


\begin{abstract}
General relativity predicts the presence of a thin, bright ring of light superimposed on the image of a black hole (the ``photon ring''), whose geometry is sensitive to the spin of the black hole.
Detecting this photon ring in observations of nuclear supermassive black holes would yield insight into gravity in the strong-field regime as well as their growth histories. 
The Black Hole Explorer (BHEX) is a mission being developed to detect the photon ring for the first time, raising the question of how precisely we can constrain spin from its observations. 
We use the method of Fisher information matrices (FIMs) along with a simple dual-cone semi-analytic emission model to forecast BHEX spin posterior widths for the primary science targets (M87* and Sgr~A*) in a variety of configurations. 
We find that BHEX can constrain the dimensionless spin of its targets to a precision of $\sigmastar \ll 0.1$ after 30 orbits, even in the presence of significant systematic errors. 
We validate our results via numerical checks for stability and robustness, as well as synthetic data cross-validation to assess handling of covariances and prior information in the FIM against a full posterior exploration. 
\end{abstract}

\maketitle

\section{Introduction}

The no-hair theorem of general relativity guarantees that the spacetime of a stationary black hole is fixed by its mass, angular momentum (spin), and charge. 
In astrophysical environments, however, surrounding plasma rapidly neutralizes any residual charge, reducing the problem to just two defining numbers \cite{1963PhRvL..11..237K, 1971PhRvL..26..331C, 1975PhRvL..34..905R, 2012LRR....15....7C, 2015CQGra..32l4006T}. 
Spin and mass together fix the geometry of the spacetime. In particular, a black hole's spin acts as an extractable energy reservoir; consequently, it plays a deciding role in whether electromagnetic jets are fueled directly by the black hole or by the accretion disk \cite{1969NCimR...1..252P, 1977MNRAS.179..433B, 1982MNRAS.199..883B}. 
Over cosmological timescales, the spin distribution of supermassive black holes serves as an enduring imprint of their assembly history \cite{1970Natur.226...64B, 2003ApJ...585L.101H, 2005ApJ...620...69V}. 
While recent gravitational-wave measurements have yielded spin constraints for stellar-mass black holes \cite[e.g.,][]{2025ApJ...993L..21A}, a measurement of spin for even a single supermassive black hole with well-characterized systematics therefore constrains both the near-horizon physics and the growth history of the population.

Measuring supermassive black hole spins traditionally relies on X-ray continuum fitting or relativistically broadened reflection spectroscopy. While these techniques have successfully constrained spins for numerous X-ray binaries and a smaller subset of active galactic nuclei \cite[AGN; ][]{1997ApJ...482L.155Z, 1989MNRAS.238..729F, 2006ApJ...652..518M}, they carry notable limitations. both are strongly dependent on assumptions about the underlying model \cite{2021ARA&A..59..117R, 2025MNRAS.544.2880M, 2024MNRAS.531..366M}. Additionally, neither approach works for the low-luminosity sources that make up the vast majority of the observable AGN population.
The Event Horizon Telescope (EHT) recently delivered groundbreaking horizon-scale mass measurements for M87* and Sgr~A* \cite{2019ApJ...875L...6E, 2022ApJ...930L..12E}. However, deriving spin from these images remains difficult, and current EHT spin constraints are tenuous and indirect. Because they require matching observations against extensive libraries of general relativistic magnetohydrodynamic (GRMHD) simulations, these estimates are limited by the uncertain plasma microphysics of those simulations \cite{2019ApJ...875L...5E, 2021ApJ...910L..13E, 2022ApJ...930L..16E, 2024ApJ...964L..26E, 2024ApJ...974..143C, 2026ApJ..1000..231B}. 
In the past few years, analyses of the lensed substructure of emission around the black hole (the ``photon ring'') show the image carries a substantial and near-universal imprint of the black hole's spin \cite{2020SciA....6.1310J, 2020PhRvD.102l4004G}. 
A number of proof-of-concept studies have established that, given access to the photon ring, a measurement of spin is in principle possible: the shape of the $n=1$ sub-ring is a function of mass, spin, and inclination, and is largely insensitive to the astrophysics of the emitting plasma \cite{2020PhRvD.102l4003G, 2020PhRvD.101h4020H, 2020ApJ...900...77F, 2021PhRvD.103j4038H, 2022ApJ...927....6B, 2025PhRvD.111j3042K, 2026ApJ..1003...63F, 2026arXiv260324722G}.

The Black Hole Explorer (BHEX) is designed to provide access to measurements of spacetime parameters through the photon ring \cite{2024SPIE13092E..2DJ, Marrone_2024}. 
By placing a single antenna in a medium Earth orbit and observing at multiple frequencies (80 GHz, 240 GHz, 320 GHz) in conjunction with a ground array, BHEX will access baselines of up to $\sim35 \ G\lambda$, where (for M87* and Sgr~A*) the direct ($n=0$) subimage is ``resolved out", and the $n=1$ photon subring appears as a periodic modulation of the visibility amplitude whose period encodes the diameter and shape of the ring \cite{2020SciA....6.1310J, 2024SPIE13092E..6QL}. 
A major goal of BHEX is to measure the spins of M87* and Sgr A* \cite{2024SPIE13092E..2DJ}. 
The exact projected precision of this measurement and its sensitivity to properties of these black holes and BHEX design choices is not obvious, for several reasons.
First, the $n=1$ ring is not the critical curve \cite{1973ApJ...183..237C}; its apparent shape is displaced and distorted by the accretion properties, optical depth, and variability of the emitting plasma, and these astrophysical effects are partially degenerate with spin \cite{2019PhRvD.100b4018G, 2020SciA....6.1310J, 2020PhRvD.102l4004G, 2022ApJ...939..107P, 2026ApJ..1003...63F}. 
Second, although long baselines ($u \gtrsim 20 \mathrm{\ G\lambda})$ preferentially isolate the photon ring by resolving out the direct emission, the absolute correlated flux density falls steeply with baseline length (falloff $\propto |u|^{-3/2}$, \citealt{2020SciA....6.1310J}) so the baselines that constrain the ring carry only a few mJy \cite{2024SPIE13092E..2DJ, 2024SPIE13092E..6QL}. 
The BHEX baselines are long enough to measure the angle-dependent diameter and brightness of the $n=1$ photon ring, but are insufficient to measure the precise radial profile \cite{2024PhRvD.110h3044J, 2023PhRvD.108f4043C}. 

In this Letter, we forecast the spin uncertainty BHEX can achieve, taking into account the interferometric coverage, anticipated error budgets, and source characteristics. 
In \autoref{sec:formalism}, we describe our methodology, considerations, and assumptions. 
In \autoref{sec:validation}, we validate our method using a combination of numerical checks, tests on existing instruments, and self-fits of simulated observations.
Finally, in \autoref{sec:discussion}, we discuss our results and consider caveats to the forecasted uncertainty. 

\section{Fisher matrix formalism and application}
\label{sec:formalism}

To forecast the ability of BHEX to constrain spin, we employ the method of Fisher information matrices \cite{1925PCPS...22..700F, 2008PhRvD..77d2001V}, which have been used in a number of astrophysical experiment forecasts \cite[e.g.,][]{1994PhRvD..49.2658C, 1999ApJ...514L..65H, 2008MNRAS.389.1375B}. 
Given a set of observables $\vec{O}$, parameters $\vec{p}$, and corresponding likelihood $\mathcal{L}$, the Fisher information matrix (FIM) is the negative expected value of the Hessian matrix of the log-likelihood function, yielding the expected curvature of the log-likelihood function around a fiducial point in the model parameter space. 
Adopting the Bayesian interpretation of the FIM \cite{2003prth.book.....J}, the diagonal element of the inverse FIM $(\mathcal{F}^{-1})_{ii}$ corresponding to a parameter $p_i$ approximates the variance of the posterior probability distribution for that parameter, assuming a flat prior, Gaussian noise, and the limit of high \SNR (or large data volume). 
Thus, by constructing the FIM for a realistic BHEX observation, we may obtain a theoretical lower bound (i.e., the Cram\'er-Rao lower bound in the frequentist interpretation of the FIM; \citealt{cramer1946mathematical, Rao1992InformationAT}) on the uncertainty $\sigmastar$ of a spin measurement made by BHEX.

To apply the FIM formalism to BHEX, we make the following reasonable assumptions. 
First, we assume Gaussian thermal noise error as the dominant data uncertainty, and that data are sufficiently high \SNR to justify the Bernstein-von Mises theorem conditions (\citealt{Vaart_1998, Rao1992InformationAT}). Next, we assume that the log-likelihood is well-approximated locally by a quadratic function, which we validate in \autoref{sec:validation} by comparing the forecasted uncertainties to self-fits performed using our chosen model. 
Additionally, the FIM is also assumed to be evaluated at a particular fiducial point; though we report only one fiducial point for consistency, we check alternative fiducial points to assess how much the forecasted uncertainty varies. 
Finally, we assume that our chosen model is representative of the source; while it is well-established that semi-analytic ray-traced models merely approximate reality, the FIM does not natively report uncertainty associated with model mis-specification. 

Our chosen model is the analytic dual-cone accretion model proposed in \citet{2024ApJ...974..143C} propagated through the \krangjl\footnote{https://dominic-chang.com/Krang.jl/stable/} ray-tracer for analytic null geodesics in the Kerr spacetime \cite{chang_2024_13936258}. 
This model describes synchrotron radiation originating from a magnetized, axisymmetric plasma channeled through twin opposing cones oriented parallel to the black hole's spin axis. 
The model is equipped with 11 parameters composing $\vec{p}$: (i) $M/D$, the mass-to-distance ratio  of the black hole, (ii) $a_*$, the dimensionless spin of the black hole, (iii) $\theta_o$, the observer polar angle, (iv) $\theta_s$, the opening angle of the cone, (v) $r_{\rm peak}$, the number density function peak radius, (vi) $p_1$, the number density function inner power-law slope, (vii) $p_2$, the number density function outer power-law slope, (viii) $\chi$, the azimuthal angle of the fluid velocity in the zero-angular-momentum-observer (ZAMO) frame, (ix) $\iota$, the angle of the magnetic field inclination in the fluid frame, (x) $\beta$, the magnitude of the fluid velocity, and (xi) $\alpha$, the spectral index of the synchrotron emission. 
The midpoint of prior ranges of the above parameters are assumed as a fiducial point for the purposes of the FIM, and, unless otherwise specified, this combination of parameters is used in visualizations.  
For a more thorough description of the model parameters and the prior ranges used, see \cite{2024ApJ...974..143C}.

We construct the observables $\vec{O}$ based on the interferometric quantities measured during a real BHEX observation. 
A natural choice might be complex visibilities $V$, as these are the most direct interferometric data product. 
However, visibilities suffer from sensitivity to site-dependent gain factors, and BHEX will have no absolute phase reference, making calibration of absolute phases challenging. 
As an alternative, we can instead compose our observables of amplitudes ($|V|$) and closure phases (the argument of the bispectrum, $\arg V_{12}V_{23}V_{31}$, see \citealt{Thompson2001, Chael2018}). 
Under this prescription, complex gains may be well-approximated to first order by a global fractional systematic error \cite{2019ApJ...875L...3E, 2019ApJ...882...23B}, allowing us to estimate the impact of poor calibration without requiring fitting dedicated gain parameters for each station in the array. 
We note that the use of amplitudes and closure phases may violate our assumption of normal errors, as amplitudes are necessarily Ricean-biased and large closure phase errors are difficult to interpret due to phase-wrapping \cite{Thompson2001, 2020ApJ...894...31B}. 
However, due to the coherence and sensitivity advantages of frequency phase transfer (FPT, \cite{2005A&A...433..897M, RD_review, RD_ngeht, Issaoun_2025}, see \appref{appendix:ngehtsim} for further details), all our measurements are in a sufficiently high-\SNR regime that the assumption of Gaussian errors is preserved for the vast majority of measurements. 
Finally, we use the \texttt{Julia} autodifferentiator \texttt{Enzyme.jl} \cite{NEURIPS2020_9332c513, 10.1145/3458817.3476165} to compute the Jacobian $J$ of our observables with respect to variations in our chosen parameters $\vec{p}$.

We generate realistic synthetic BHEX observations of M87* and Sgr A* using the \texttt{ngEHTsim}\footnote{https://github.com/Smithsonian/ngehtsim}  code \citep{Pesce_ngehtsim_2023, Pesce_ngehtsim_2024}, which models both weather and instrumental effects, re-sampling our semi-analytic source model onto the detection patterns and thermal noise properties of existing synthetic data sets built for ongoing BHEX imaging studies (Zeng et al., in preparation). 
Each source is observed once every three days over a three-month campaign by a modest ground array together with BHEX in a $\sim$12 hour polar orbit whose ascending node is chosen to place the orbital plane perpendicular to the line of sight to M87*, which also yields a nearly north--south baseline track toward Sgr A* that mitigates its predominantly east--west scatter broadening \citep{Tamar_2025}. 
The synthetic data incorporate simultaneous multifrequency observations with FPT from a $\sim$80 GHz band to the tunable high band, set to 240 GHz for M87* and 320 GHz for Sgr A* \citep{RD_review, RD_ngeht, Marrone_2024, Tong_2024, Issaoun_2025}. Full details of the array composition, observing windows, and FPT thresholds are given in \appref{appendix:ngehtsim}.

The covariance matrix is a critical component of the FIM, as it incorporates observable uncertainties and correlations directly into the likelihood calculation. 
In BHEX observations, the primary noise source on baselines sensitive to the photon ring is thermal noise \cite{Marrone_2024}. 
Because thermal noise originates at individual antenna stations (which are shared across multiple baselines and closure triangles) it introduces complex, non-trivial correlations across the data. To construct our covariance matrix $C$, we propagate per-measurement uncertainties generated in the simulated observations by \texttt{ngEHTsim} \cite{Pesce_ngehtsim_2023} into the diagonal of $C$. 
To approximate the effect of complex gains, a fractional (by amplitude) systematic error $f_{\rm sys}$ is added in quadrature to the existing amplitude uncertainties (note that closure phases are gain-insensitive and therefore do not require this correction; \cite{Thompson2001, Chael2018, 2020ApJ...894...31B}). 
Off-diagonal covariances are well-understood and are generated directly following Blackburn \cite{2020ApJ...894...31B}. 

The effects of scattering for Sgr~A* fall into two categories \citep{Narayan_1992, Johnson_Narayan_2016}. The first, diffractive scattering, results in angular broadening described by a time-independent kernel. If the kernel is known, as is the case for Sgr~A* from observations at longer wavelengths \citep{Psaltis_2018, 2018ApJ...865..104J}, these effects can be directly inverted by dividing the visibilities by the scattering kernel \citep{Fish_2014}; the effect is then to amplify thermal noise on long baselines. The second, refractive scattering, results in large-scale deformations of the image that correspond to stochastic noise added to long-baseline measurements and have zero mean when averaged over many scattering timescales. In the present analysis, we focus on the information contained in the average image; we account for the effects of diffractive scattering for Sgr~A* but not refractive scattering. Because the effects of scattering for observations of M87* are approximately 1000 times weaker than for Sgr~A*, they are negligible for submillimeter observations, even those with BHEX.

Finally, we incorporate information known \textit{a priori} into the FIM additively via a prior matrix $\mathcal{F}_{\rm prior}$.
For all parameters $p_i$ (excluding $M/D$ for Sgr~A*), we assume a log-prior density of constant curvature matching the standard deviation expected from a flat prior ($(\max p_i - \min p_i)/\sqrt{12}$, where max and min refer to the upper and lower bounds of the prior, respectively). 
This choice enforces that, in the limit of no data-related Fisher information, the total Fisher information gives the standard deviation expected from a flat, bounded prior.
For $M/D$ of Sgr A*, we assume a prior uncertainty of $\approx1\%$ based on measurements from GRAVITY \cite{2021A&A...647A..59G} and UCLA \cite{2019Sci...365..664D}, but vary this prior between 0.0\% and 100\% as part of our validation (see \autoref{sec:validation}).

To compute the forecasted spin uncertainty, we construct the total FIM $\mathcal{F}_{\rm tot}$ by adding the data and prior FIMs as described above,
\begin{align}
    \mathcal{F}_{\rm tot} = \mathcal{F}_{\rm data} + \mathcal{F}_{\rm prior},
\end{align}
where $\mathcal{F}_{\rm data}$ is constructed via the standard matrix method,
\begin{align}
    \mathcal{F}_{\rm data} = J^{\intercal} C^{-1} J. 
\end{align}
The forecasted spin uncertainty $\sigma_{a_*}$ is then the square root of the $p_a$ entry along the diagonal of the inverted total FIM, i.e., 
\begin{align}
    \sigma_{a_*} = \sqrt{[\mathcal{F}_{\rm tot}^{-1}]_{a_* a_*}}.
\end{align}
We compute $\sigma_{a_*}$ at fiducial points corresponding to every point in the Kerr ($a_*, \theta_o$) parameter space at three combinations of the remaining parameters (excluding $M/D$) to ensure non-sensitivity to the choice of fiducial point (expanded on in \autoref{sec:validation}). 

\begin{figure*}
    \centering
    \includegraphics[width=0.45\linewidth]{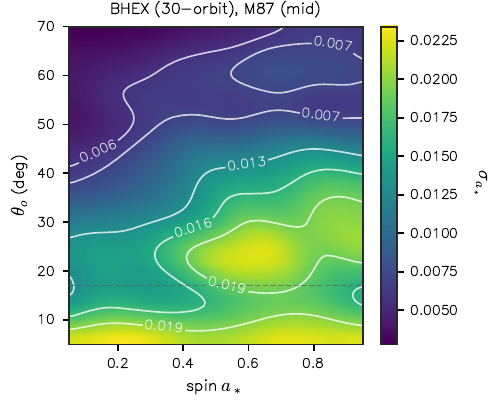}
    \includegraphics[width=0.45\linewidth]{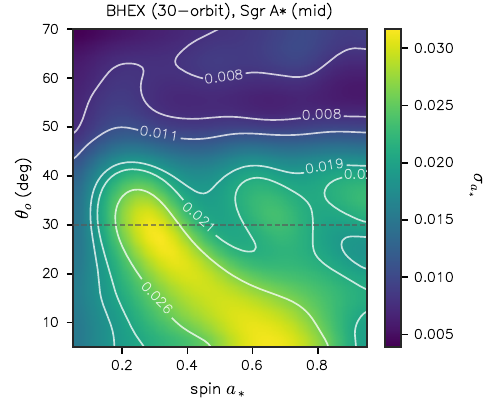}
    \caption{Heatmap of the forecasted BHEX inference capability $\sigmastar$ as a function of dimensionless spin $a_*$ and inclination $\theta_o$. We compute the forecasted uncertainty via $\sigmastar=\sqrt{(F^{-1})_{a_* a_*}}$ at each combination of spin and observer inclination for fiducial plasma physics parameters in the middle of the prior range. Horizontal dashed lines denote the nominal fiducial observer inclinations ($\theta_o \approx 17^\circ$ for M87* and $\theta_o \approx 30^\circ$ for Sgr A*). The displayed results are computed from a FIM incorporating a fractional systematic error $\fracsys$ applied in quadrature to amplitudes only. The reported uncertainty is sufficiently precise ($\sigmastar \ll 0.1$) for both sources at every point in the Kerr parameter space.}
    \label{fig:sigma_mapp_kerr}
\end{figure*}

The results of this computation are displayed in \autoref{fig:sigma_mapp_kerr} for both M87* and Sgr A*.  Despite the differences in coverage alignment and \textit{a priori} $M/D$ constraints, both sources yield $\sigmastar\lesssim0.03$ for the entire Kerr parameter space. At fiducial values ($a_\ast=0.5$, $\theta_{\rm o, \ M87^*}=17^\circ$, \citealt{1997AAS...19110410B, 2018ApJ...855..128W, 2019ApJ...875L...5E};  $\theta_{\rm o, \ Sgr \ A^*}=30^\circ$, \citealt{2022ApJ...930L..16E}), we forecast BHEX spin uncertainties of $\sigmastar\approx0.019$ for M87* and $\sigmastar\approx0.021$ for Sgr A*. We forecast worse (i.e., higher) uncertainties generally at near-face-on inclinations and low spins, consistent with expectations from the theory of the critical curve \cite[see, e.g.,][]{2020ApJ...900...77F}, as well as previous analyses in the literature \cite[e.g.,][]{2022ApJ...939..107P, 2024ApJ...974..143C, 2025PhRvD.111j3042K,2025ApJ...983..185F,2026ApJ..1003...63F}. 
The heatmap for Sgr A* (and M87* to a lesser extent) at low spins displays a minor reversal from this trend, with smaller uncertainties at $\theta_o \lesssim20^\circ$ than at $\theta_o\sim30^\circ$. This behavior may be explained by the recent discovery that contours of key image observables in $(a_\ast, \theta_{\rm o})$ space become nearly orthogonal in this near-face-on regime, providing tight parameter constraints that do not persist for the same observables at moderate or high inclinations \cite{2026ApJ..1003...63F}.

\section{Validation}
\label{sec:validation}

FIMs can often yield unrealistically optimistic forecasts. The most common points of failure for FIMs are well-documented in the literature and have standard diagnostics and validations; for the following, we draw from the treatment of Vallisneri \cite{2008PhRvD..77d2001V}. 
Here, we validate the FIM results from \autoref{sec:formalism} by investigating the most-common and most-likely points of failure present in our analysis. 
The points of failure can be grouped into three categories: (i) failures of assumptions, (ii) insufficient exploration of framework choices, and (iii) over-interpretation of the theoretical performance upper bound.

We begin by examining whether the key assumptions intrinsic to the FIM remain valid. 
The quadratic approximation of the log-likelihood assumes that the behavior of the likelihood around the fiducial point is roughly Gaussian (and therefore symmetric by construction). 
However, when parameters are bounded (as $a_\ast$ and $\theta_o$ are; $a_\ast \in [0, 1)$ and $\theta_o \in [0, 90^\circ]$), this assumption may be invalidated at the boundary, leading to an incorrect uncertainty forecast. 
Indeed, the fiducial M87* inclination is $\theta_o\approx17^\circ$, quite close to the face-on inclination boundary, and previous analyses suggest either M87* or Sgr A* could have close-to-maximal spin \cite{2012MNRAS.423.3083M, 2019ApJ...880L..26N, 2024MNRAS.527..428D}. There are several established methods to combat boundary misbehavior, such as transforming to a truncated multivariate Gaussian or using the constrained Cram\'er-Rao lower bound when the $\pm1\sigma$ estimate exceeds the boundary \cite{Nitzan2018}. Here, we find the uncertainty reported is sufficiently low that adverse effects are limited to a narrow region (within $3\sigmastar$ of the boundary) at the edges of the parameter space. 

We next evaluate the numerical stability of the FIM inversion, which is equivalent to assessing whether parameter degeneracies render the fit under-determined or ill-conditioned \cite[see e.g.,][]{2021arXiv210100298B}. 
Although the number of observables (individual amplitudes and closure phases) vastly outnumbers the number of model parameters, high correlations among observables can render the FIM ill-conditioned or near-singular. 
To assess inversion stability beyond simple determinant criteria, we compute the condition number $\kappa(\mathcal{F}_{\rm tot}) = \lambda_{\max}/\lambda_{\min}$ (where $\{\lambda_j\}$ are the eigenvalues of the FIM) following Vallisneri \cite{2008PhRvD..77d2001V} across all parameter configurations. 
We find that $\kappa(\mathcal{F}_{\rm tot}) \sim 10^6$ for all calculations in this work, and confirm that the matrix inversion is stable to small random perturbations of the  FIM elements. 

\begin{figure}
    \centering
    \includegraphics[width=1\linewidth]{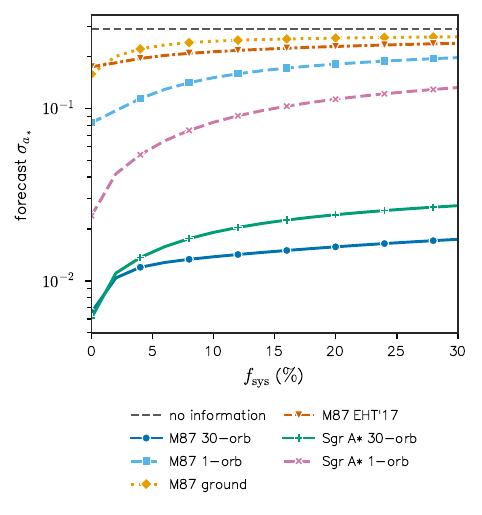}
    \caption{Variation of the forecasted spin uncertainty ($\sigmastar$) as a function of injected systematic error fraction ($\fracsys$) evaluated between $0\%$ and $30\%$. In the 30-orbit BHEX configuration, both M87* and Sgr A* are precisely measured ($\sigmastar \ll 0.1$) across all tested values of $\fracsys$. For the 1-orbit configuration, Sgr A* achieves higher precision, whereas M87* satisfies $\sigmastar < 0.1$ only for low systematic error levels ($\fracsys \lesssim 5\%$). For comparison, ground-only configurations (single-night EHT 2017 and the BHEX ground array) show negligible ability to constrain spin, approaching the uninformative uniform prior limit of $\sigmastar = 1/\sqrt{12}$ (horizontal dashed line).}
\label{fig:fsys_variation}
    \label{fig:fsys}
\end{figure}

Having validated that the major assumptions of the FIM hold in the analysis, we next examine whether our analysis is sensitive to choices made in the construction of our framework. 
A significant choice made early on is to approximate complex gain effects via a fractional systematic error $\fracsys$, motivated by Blackburn \cite{2019ApJ...882...23B}. 
However, the magnitude of this error is largely dependent on the characteristics of the yet-to-launch BHEX spacecraft and component ground array. 
To investigate the impact of this choice, we vary the incorporated $\fracsys$ between 0\% and an unrealistically high 30\%. 
The results of this experiment are shown in \autoref{fig:fsys}. 
We find that the impact on forecasted $\sigmastar$ largely plateaus above $\fracsys \sim 10\textrm{--}15\%$. 
Even with unrealistically high choices of $\fracsys$, the BHEX 30-orbit configuration achieves $\sigmastar\sim0.02$ for both M87* and Sgr A*.

Another important methodological choice is the selection of the fiducial parameter point. 
While we systematically explore the entire Kerr parameter space ($a_\ast, \theta_o$), we set the remaining parameters' fiducial values to the midpoints of their priors. 
We investigate the impact of the fiducial point on the forecasted uncertainties by evaluating the FIM across several parameter locations that evenly sample the prior space. 
We find that the choice of fiducial parameter point can affect the estimate of $\sigmastar$ by up to a factor of $\approx2$ in the BHEX 30-orbit configuration and up to $\approx30\%$ in the BHEX 1-orbit configuration. 
These shifts ($\sigmastar{}_{, \ 30}\sim0.01\to0.02$ and $\sigmastar{}_{, \ 1}\sim0.14\to0.19$) do not materially change the interpretation of the results.

\begin{figure}
    \centering
    \includegraphics[width=1\linewidth]{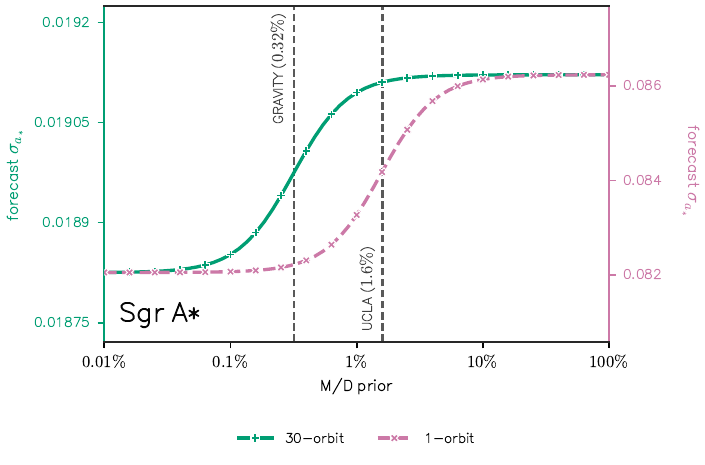}
    \caption{Sensitivity of the forecasted BHEX spin uncertainty to the choice of $M/D$ prior width. We vary the $M/D$ entry in the Fisher prior matrix $\mathcal{F}_{\rm prior}$ between $0.0\%$ and $100\%$ and recompute the predicted precision $\sigmastar$. We find that the effect is negligible ($\lesssim5\%$ for the 1-orbit configuration, pink, and $\lesssim 3\%$ for the 30-orbit configuration, green). We show the constraints from UCLA/Keck \cite[$\approx1.6\%$;][]{2019Sci...365..664D} and GRAVITY \cite[$0.32\%$;][]{2021A&A...647A..59G} as dashed lines.}
    \label{fig:sgra_md}
\end{figure}

The last major framework choice is the inclusion of an \textit{a priori} constraint on the $M/D$ of Sgr A*, which could plausibly have a significant impact on the ability of BHEX to constrain spin.
Multiple analyses have found that, when $M/D$ is known, the size of the photon ring in the black hole image is an important discriminant for spin \cite{2020ApJ...900...77F, 2025ApJ...983..185F, 2026ApJ..1003...63F, 2025PhRvD.111j3042K}. 
The $M/D$ of Sgr A* is tightly constrained by previous analyses to a precision of $\lesssim2\%$ \cite{2019Sci...365..664D, 2022A&A...657L..12G}. 
To investigate the impact of this prior on our forecast, we vary the width of the $M/D$ prior between $0.0\%$ and $100\%$ and calculate the resulting forecasted $\sigmastar$ for both the 30-orbit and 1-orbit configurations. 
The results of this computation are shown in \autoref{fig:sgra_md}. 
We find that increasing the $M/D$ prior by an order of magnitude from $0.1\%$ to $1\%$ increases the forecasted $\sigmastar$ by $\sim2\%$ in the 1-orbit configuration and $\sim 1\%$ in the 30-orbit configuration, a negligible effect. 

\begin{figure}
    \centering
    \includegraphics[width=1\linewidth]{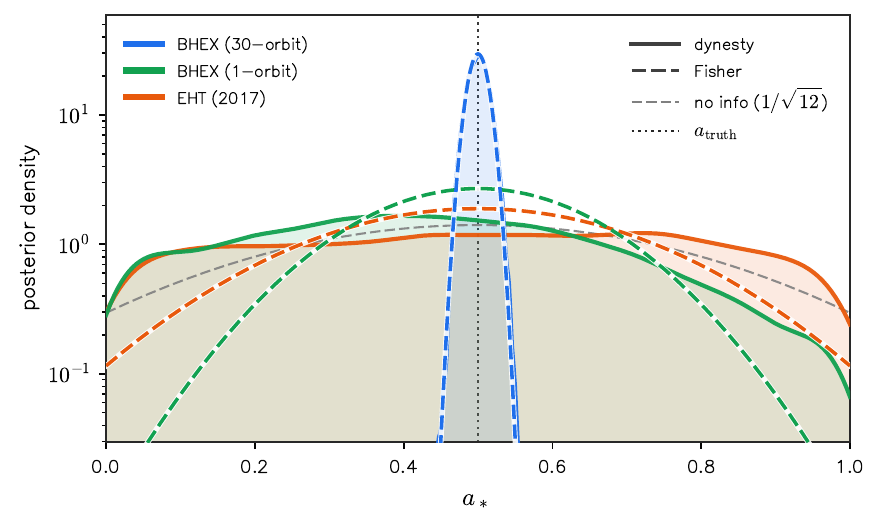}
    \caption{Validation of the forecasted uncertainty against self-fit posterior widths. We perform self-fits using a dual-cone model ray traced with \krangjl and compare the resulting $a_*$ posterior widths to the widths predicted by the FIM analysis. The Fisher forecast matches the self-fit posterior standard deviation to within $\lesssim 3\%$ for the BHEX 30-orbit configuration, while accurately capturing the EHT 2017 array's inability to constrain spin by predicting $\sigmastar \approx 1/\sqrt{12}$, which corresponds to a nearly flat, uninformative posterior. The posteriors retain their normalization to unity and are shown on a log scale to facilitate comparison.}
    \label{fig:posterior}
\end{figure}

Finally, having considered assumptions and framework choices, we perform several sanity checks to avoid over-interpreting the forecast. 
In particular, even with the most careful treatment of covariances and assumptions, the FIM fundamentally represents a lower bound on parameter uncertainties, achieved only in the limit of high-$\SNR$. 
To assess the degree of optimism in our Fisher-based forecast, we perform an end-to-end test of parameter recovery using the same model we have described in this Letter applied to a simulated BHEX observation. 
Given the simulated BHEX (1-orbit and 30-orbit) or EHT 2017 dataset (amplitudes and closure phases only) of a dual-cone model at the fiducial parameter set for M87*, we incorporate $\fracsys\sim10\%$ and perform a self-fit of the dual-cone model to the simulated observation. 
Parameter space exploration is performed with the \texttt{dynesty}\footnote{https://dynesty.readthedocs.io/en/v3.1.0/} dynamic nested sampler \cite{2020MNRAS.493.3132S} and continued until convergence is achieved. 
The resulting posteriors and predicted posterior widths are shown in \autoref{fig:posterior}.
We find that the FIM predicts the posterior width of the self-fit to $\lesssim3\%$ for the BHEX 30-orbit configuration, and correctly identifies (via $\sigmastar \sim 1/\sqrt{12}$) that the EHT 2017 array cannot constrain spin (yielding an uninformative, almost uniform posterior). 
The accurate prediction of the self-fit posterior width indicates that our forecast is not significantly overestimating the ability of BHEX to perform the measurement at the level reported.

\section{Discussion}
\label{sec:discussion}
In this Letter, we have established that BHEX is capable of achieving its goal of a precise measurement for both of its primary targets, M87* and Sgr A*. 
The anticipated performance maximum of $\sigmastar\sim 0.02$ is robust against common FIM pitfalls, framework choices, and several major sources of observational uncertainties. 
However, we emphasize that this result has several key caveats, which we discuss below.

First, the long dynamical time scale of M87* makes the source relatively straightforward to image \citep[the mass measurement from][implies a gravitational time scale of ${\sim}~8~\text{hours}$]{2019ApJ...875L...6E}.
However, the long dynamical time scales also mean that correlated structures could persist over the span of a single BHEX observation, potentially bias parameter inference \citep[][]{Conroy_2023, Chang_2025}.
Sgr A* also has several additional sources of uncertainty that will complicate spin recovery: scattering and variability. 
Owing to a significant amount of turbulent, ionized plasma along the line-of-sight to the Galactic Center, scattering of the interstellar medium significantly contaminates the interferometric signal, depressing the signature of the $n=1$ photon ring on long baselines and altering the features of the image in a complex, non-invertible manner \cite{2018ApJ...865..104J, 2019ApJ...870....6Z, 2019ApJ...871...30I}. 
While we have partially addressed this source of contamination in this work, a full treatment of diffractive and refractive scattering is required to assess the impact on the measurement. 
Additionally, Sgr~A* has a dynamical timescale a factor of $\sim10^3$ shorter than M87* (the gravitational time scale of Sgr A* is ${\sim}~20s$) resulting in high variability on timescales far shorter than the aperture synthesis timescale \cite{Macquart_2006, 2022ApJ...930L..15E, 2022ApJ...930L..14E}. 
Both of these effects are likely to increase the parameter uncertainty for 1-orbit observations over the forecast but are partially mitigated by averaging over the full 30-orbit observation to remove transient features. Additionally, despite being a confounding factor, variability itself can provide information about the angular momentum of the source \cite[see e.g.,][]{2021PhRvD.103j4038H, Conroy_2023,Zhou:2024dbc}.

Second, any FIM analysis is inextricably dependent on the model specification, and the assumption that the chosen model reflects reality to the extent probed by the observations. 
If the model is overly simplistic or deviates significantly from reality, the model mis-specification introduces systematic biases into the recovered parameters to which the FIM (evaluating only local log likelihood curvature) is completely insensitive.
While this concern is relevant here (as current models of black hole environments remain incomplete, and the error budget of BHEX has not been fully characterized), we have several reasons to expect our results to remain robust.
The dual-cone model utilized in this Letter has demonstrated success in fitting legacy interferometric data, parameterizing both the background spacetime geometry and the primary emitting components (accretion disk and jet). 
Additionally, our validation of the forecast's robustness against significant systematic errors (\autoref{sec:validation}) demonstrates that the framework can accommodate substantial unmodeled noise while still achieving a spin precision of $\sigmastar < 0.1$. 
Finally, as a semi-analytic model, the dual-cone model offers a performance-friendly Jacobian computation that cannot be matched by more accurate but vastly slower GRMHD simulations. 

The most important caveat of any FIM analysis is that the reported uncertainty is merely a lower bound on the precision the measurement could, in principle, achieve. 
Such a statement makes no guarantee that there exists or will exist a method that can actually saturate this lower bound. 
For a novel experiment such as the BHEX spin measurement, one can argue that the reported forecast should not be considered realized until multiple, independent methods report the ability to achieve the target precision on the most realistic simulated dataset possible. 
However, both this Letter and the literature present several meaningful rebuttals to this concern. 
First, our own validation of the posterior width (\autoref{sec:validation}) presents a concrete proof-of-concept method which achieves the forecasted uncertainty on a realistic, if ideal, simulated observation. 
Additionally, there is an active and growing body of literature demonstrating multiple promising procedures to achieve a BHEX spin measurement in independent ways \cite{2020PhRvD.102l4003G, 2020PhRvD.101h4020H, 2020ApJ...900...77F, 2021PhRvD.103j4038H, 2022ApJ...939..107P, 2022ApJ...927....6B, 2024ApJ...974..143C, 2025PhRvD.111j3042K, 2025ApJ...983..185F, 2026ApJ..1003...63F, 2026arXiv260324722G}, including using constraints available from data not included in this analysis (e.g., polarization, multi-band data, variability). 
Future work on developing BHEX spin inference capabilities will focus on bringing these multiple methods to maturity in order to realize unprecedentedly precise and robust black hole spin measurements.

\acknowledgements
We thank Peter Galison, Alex Lupsasca, Angelo Ricarte, and Paul Tiede for their insightful discussions.
We acknowledge financial support from the National Science Foundation (AST-2307887). 
This Letter is funded in part by the Gordon and Betty Moore Foundation (Grants \#13526 and \#12987). It was also made possible through the support of a grant from the John Templeton Foundation (Grant \#63445).  The opinions expressed in this publication are those of the author(s) and do not necessarily reflect the views of these Foundations.

\appendix
\section{Generation of simulated observations}
\label{appendix:ngehtsim}
We simulate realistic synthetic VLBI data using the \texttt{ngEHTsim} code, which incorporates both weather and instrument modeling to generate interferometric observations \citep{Pesce_ngehtsim_2023, Pesce_ngehtsim_2024}. In particular, we re-use existing synthetic data sets generated in service of ongoing BHEX imaging studies of M87* and Sgr A* (Zeng et al, in preparation). These synthetic data sets assume that M87* and Sgr A* are each observed by BHEX alongside a modest ground array once every three days over a three month campaign. Though these data sets were generated with an on-sky image model of a time-varying general relativistic magnetohydrodynamical simulation, we re-use the forward-modeled VLBI detections and resulting thermal noise properties and instead re-sample the full campaign of data on our comparatively simple semi-analytic model. As a result, we are able to capture more of the stochasticity of realistic detection prospects, incorporating variability that is not captured innately by our source model.

In our data projections, BHEX is assumed to reside in a ~12 hour polar orbit with a right ascension of the ascending node of 277 degrees, chosen so that the orbital plane is perpendicular to the line of sight to M87*. This orientation maximizes the angular sampling of the visibility response to the M87* photon ring, providing excellent two-dimensional constraint of its size, shape, and relative shift. Meanwhile, due to M87* and Sgr A* having a nearly 90 degree separation, this orbit projects to a nearly north-south aligned ellipse as seen from Sgr A*, ideal for avoiding the worst of the Sgr A* scatter-broadening, which is predominantly east-west \citep{Tamar_2025}. While the reduced angular coverage of long baselines reduces the intrinsic spin sensitivity of the coverage, the much-better-known mass-to-distance ratio of Sgr A* makes quasi-one-dimensional Fourier sampling a tractable measurement pathway.

The \texttt{ngEHTsim} synthetic data model incorporates simultaneous multifrequency observation with an emulation of frequency phase transfer \citep[or ``FPT'', described in ][]{RD_review, RD_ngeht}, a technique crucial to projected BHEX observations of M87* in which strong detections at a low $\sim80$ GHz observing band are used to steer the phase of the higher tunable band center, tuned to 240 GHz for our M87* observations and 320 GHz for our Sgr A* observations \citep{Marrone_2024, Tong_2024}. This technique was recently demonstrated in the 80-240 GHz transfer on terrestrial baselines by \citet{Issaoun_2025}. \texttt{ngEHTsim} emulates FPT by defining a threshold S/N in the low-band beyond which the high-band data will always be recorded regardless of high-band S/N; this threshold is set to thrice the frequency ratio (that is, 9 for 80-240 GHz and 12 for 80-320 GHz) in our synthetic data. Scatter broadening at 80 GHz makes this largely irrelevant for Sgr A*, which is directly detected at 320 GHz thanks to its larger compact flux density.


The M87* observing campaign is assumed to contain the Large Millimeter Telescope (LMT), Submillimeter Array (SMA), James Clerk Maxwell Telescope (JCMT), the IRAM 30-m Telescope (IRAM), the Greenland Telescope (GLT), Kitt Peak Telescope (KP), the Green Bank Telescope (GBT), and the Atacama Large Millimeter/Sub-millimeter Array (ALMA). Of these, all are assumed to operate with simultaneous dual-band observation except for the SMA (240 GHz only), GLT (80 GHz only), and ALMA (240 GHz only). The observations are generated between January 1st and March 31st; ALMA joins only in March, to emulate typical winter closure schedules.

For Sgr A*, the situation is similar, albeit GLT and GBT do not participate, while the African Millimeter Telescope joins the array. All high-band sites operate at 320 GHz instead of 240 GHz, and the observations take place between June 1st and August 31st.

\bibliography{ref}

\end{document}